\documentstyle[psfig,namedreferences,amssymb]{spacekap}

\runningtitle{THE UNIVERSITY OF DURHAM MARK 6 GAMMA RAY TELESCOPE}

\runningauthor{P.~Armstrong ET AL.}

\def\ga{\mathrel{\mathchoice {\vcenter{\offinterlineskip\halign{\hfil
$\displaystyle##$\hfil\cr>\cr\sim\cr}}}
{\vcenter{\offinterlineskip\halign{\hfil$\textstyle##$\hfil\cr
>\cr\sim\cr}}}
{\vcenter{\offinterlineskip\halign{\hfil$\scriptstyle##$\hfil\cr
>\cr\sim\cr}}}
{\vcenter{\offinterlineskip\halign{\hfil$\scriptscriptstyle##$\hfil\cr
>\cr\sim\cr}}}}}

\def\expastro{\em Exp. Astron.\rm}

\def\jphysg{\em J. Phys. G. Nucl. Particle Phys.\rm}
\def\astropart{\em Astroparticle Phys.\rm}

\def\apj{\em Astrophys. J.\rm}

\begin{opening}

\title{THE UNIVERSITY OF DURHAM MARK 6 GAMMA RAY TELESCOPE}

\author{P. \surname{Armstrong}} 
\author{P. M. \surname{Chadwick}} 
\author{P. J. \surname{Cottle}} 
\author{J. E. \surname{Dickinson}} 
\author{M. R. \surname{Dickinson}} 
\author{N. A. \surname{Dipper}} 
\author{S. E. \surname{Hilton}} 
\author{W. \surname{Hogg}} 
\author{J. \surname{Holder}}
\author{T. R. \surname{Kendall}} 
\author{T. J. L. \surname{McComb}} 
\author{C. M. \surname{Moore}} 
\author{K. J. \surname{Orford}} 
\author{S. M. \surname{Rayner}} 
\author{I. D. \surname{Roberts}} 
\author{M. D. \surname{Roberts}} 
\author{M. \surname{Robertshaw}} 
\author{S. E. \surname{Shaw}} 
\author{K. \surname{Tindale}} 
\author{S. P. \surname{Tummey}} 
\author{K. E. \surname{Turver}}

\institute{Department of Physics, Rochester Building, Science
Laboratories,\\ University of Durham, South Road, Durham DH1\ 3LE,
United Kingdom}

\date{Accepted for publication in Experimental Astronomy, March 1998}

\end{opening}

\begin{ao}

Prof. K. E. Turver, Department of Physics, Rochester Building, Science
Laboratories, University of Durham, South Road, Durham DH1\ 3LE,
United Kingdom. e-mail k.e.turver@dur.ac.uk

\end{ao}

\begin{document}

\begin{abstract}The design, construction and operation of the University
of Durham ground-based gamma ray telescope is discussed. The telescope
has been designed to detect gamma rays in the $\le 200$ GeV region and
to achieve good discrimination between gamma ray and hadron initiated
showers in the higher energy region ($\ga 300$ GeV). The telescope was
commissioned in 1995 and a description of its design and operation is
presented, together with a verification of the telescope's performance.

\end{abstract}

\keywords{gamma ray astronomy, gamma ray telescopes, atmospheric
\v{C}erenkov technique} 

\section{Introduction} 

\subsection{Ground-based Gamma Ray Astronomy}

Very high energy (VHE) gamma ray astronomy using the ground-based
atmospheric \v{C}erenkov technique has conventionally covered energies
in the range 300 GeV to 30 TeV. In recent years it has been firmly
established as a fruitful area for astronomical investigation (see, for
example, \citeauthor{jphysrev}, \citeyear{jphysrev},
\citeauthor{Croninetal}, \citeyear{Croninetal}, \citeauthor{schubetal},
\citeyear{schubetal}).

Ground-based gamma ray astronomy in the VHE region makes use of
atmospheric cascading, where a gamma ray strikes the top of the
atmosphere and initiates a cascade of electrons and photons. The
electrons in this cascade, which are above the \v{C}erenkov threshold in
air, produce a flash of optical radiation via the \v{C}erenkov process.
This flash of light, which lasts a few ns, penetrates to ground level
where it is spread over an area of about $10000{\rm~m}^{2}$. In this
way, a ground-based detector of limited size can have an effective
collecting area approaching the extent of the \v{C}erenkov flash, and so
the intrinsically low VHE gamma ray flux ($10^{-10} {\rm~cm}^{-2}
{\rm~s}^{-1}$ for a typical source at an energy threshold of 300 GeV)
can be detected at a reasonable rate. As an example, the Crab has a VHE
$\gamma$-ray flux of $1.3 \times 10^{-10} {\rm~cm}^{-2} {\rm~s}^{-1}$
for energies $\geq 300$ GeV, leading to a detection rate of $\sim1$
gamma per minute with a typical ground-based \v{C}erenkov telescope.

The major difficulty in applying this technique successfully has been
the presence of a high rate of background of flashes due to the
spatially and temporally isotropic charged cosmic rays. Cosmic rays
(mainly protons) of similar energy to the gamma rays of interest
interact and produce cascades and \v{C}erenkov flashes which are similar
to those from gamma ray initiated cascades. For a telescope with an
energy threshold of hundreds of GeV and an aperture of 1 sq degree, a
typical gamma ray to proton flux ratio is about 0.03, and so the effects
of the background contamination are formidable. 

First-generation \v{C}erenkov telescopes addressed the problem of the
proton background by matching the telescope aperture to the angular size
of the gamma ray flash, a typical example being the University of Durham
Mark 3 telescope \cite{mark3paper}. Many of the potential astrophysical
sources of VHE gamma rays would be expected to produce modulated signals
({\em e.g.} pulsars, X-ray binaries, cataclysmic variables). Thus,
techniques based on well established methods of phase sensitive
detection were used to enhance the detection of the gamma rays against
the randomly arriving background \cite{gibsonetal}.

There are, however, differences in the detailed structure of the
detected \v{C}erenkov flash due to two facts: the proton background is
isotropic while the gamma rays emanate from point sources, and the
physics of the cascade process is different in detail for hadronic and
electromagnetic initiated showers. A major breakthrough in
discrimination between the \v{C}erenkov signal due to gamma rays and and
that due to the proton background was the successful application of
imaging (\citeauthor{Hillas1985}, \citeyear{Hillas1985},
\citeauthor{Weekesetal}, \citeyear{Weekesetal}, \citeauthor{Cawleyetal},
\citeyear{Cawleyetal}, \citeauthor{fegan}, \citeyear{fegan}). Here an
array of PMTs is employed at the focus of the telescope to enable the
shape and orientation of the \v{C}erenkov flash to be measured. On the
basis of this information, it is possible to reject more than 99\% of
the incident protons while retaining over 50\% of the gamma rays
detected (\citeauthor{Vacetal1991}, \citeyear{Vacetal1991},
\citeauthor{Reynoldsetal}, \citeyear{Reynoldsetal}). This technique has
formed the basis for many new telescopes.

\subsection{The University of Durham Mark 6 Telescope}

The University of Durham Mark 6 telescope was constructed with the
specific aim of moving to energies towards the upper end of
space-accessed energies (30 GeV for the EGRET instrument onboard the
{\em Compton Gamma Ray Observatory}). There are two clear requirements
in fulfilling such an aim; large flux collectors combined with efficient
light sensors to detect the smaller light yields, and a method of
enhancing the gamma ray signal.

The first requirement has been met by constructing $126~{\rm m}^2$ of
mirror which, combined with efficient PMTs and the fast coincidence
technique, gives a capability to detect photon densities of $\sim5~{\rm
m}^{-2}$. The second requirement is met in two ways. The fluctuations of
the position and brightness of the \v{C}erenkov light produced in $3
\times 42~{\rm m}^{2}$ collectors will discriminate against cosmic rays
in favour of gamma rays in the event selection by the telescope.

In addition, at low energies ($E \leq 200$ GeV) the problem of enhancing
the signal to noise ratio may be assisted by nature as the contribution
from the charged particle background will have reduced due to an effect
first noted by \citeauthor{tedtrev} \shortcite{tedtrev}. At these
energies, the efficiency of \v{C}erenkov light production in charged
particle induced cascades decreases; for instance, an average 30 GeV
proton cascade produces less than 1\% of the light in a gamma ray
cascade of similar energy. 

At higher energies ($E > 300$ GeV) the second requirement is met by
equipping the telescope with a 109-element camera so that imaging
studies of the detected \v{C}erenkov flash may be employed.

\begin{figure}
\centerline{\psfig{file=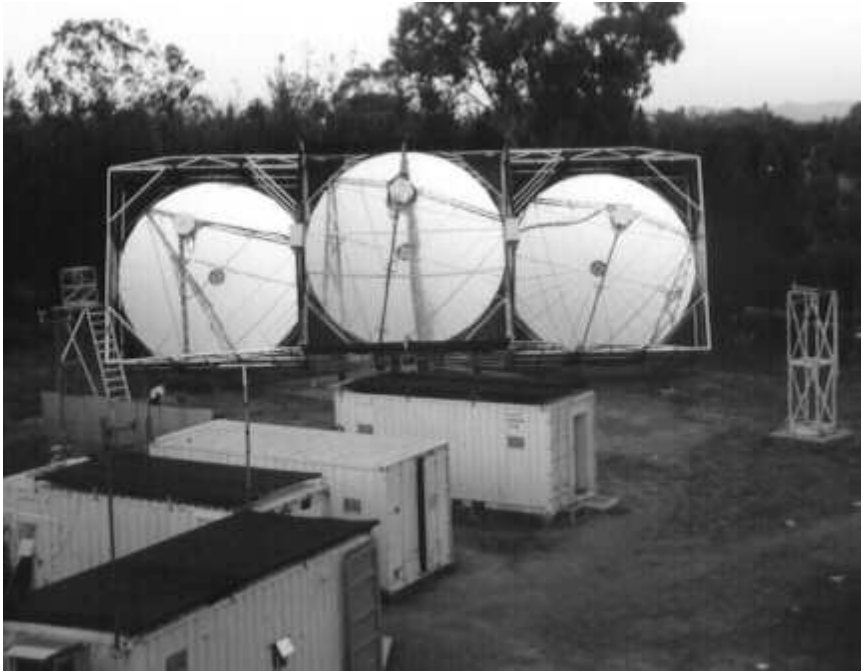,height=10cm}}

\caption{The University of Durham Mark 6 gamma ray telescope deployed at
Narrabri, N.S.W.}\label{mk6photo}

\end{figure}

A large collector area is necessary to enable the small \v{C}erenkov
flash to be detected above the sky-induced detector noise. For a given
mirror area, the detector energy threshold is determined by the smallest
\v{C}erenkov flash that can be distinguished from accidental events due
to night sky noise. For a fixed total mirror area, it is possible to
deploy the mirror in a number of ways; for example as a single
collector, as multiple collectors operated in coincidence on a single
mount or as multiple telescopes separated by $\sim 50 \rm~m$. 

The Mark 6 telescope consists of three $42 {\rm~m}^2$ parabolic flux
collectors mounted on a single alt-azimuth mount (see
Figure~\ref{mk6photo}). By operating a multiple reflector telescope and
using fast coincidence techniques, it is possible, for a fixed (small)
rate of accidental events, to run at a higher detector gain and so
achieve a lower energy threshold than either of the other options
(\citeauthor{mark3paper}, \citeyear{mark3paper}, \citeauthor{pathill},
\citeyear{pathill}). Our previous experience has shown that a triple
coincidence system is optimum for maximising the sensitivity and
achieving a low energy threshold \cite{mark3paper}. Further improvement
may be added by requiring a spatial correlation between the three
signals recorded in the three flux collectors in a narrow time window.

It is important to note that such a triple-mirror fast coincidence
system removes the background due to the effects of local muons in the
atmosphere and in the PMTs.

Another advantage of a triple coincidence system is that the two
additional samples of Cerenkov light taken by two well-separated
detectors strongly favour triggering by low energy gamma rays and allow
additional proton rejection on the basis of fluctuation measures when
analysing data in the higher energy region.

This paper describes in detail the design, construction and operation of
the Durham University Mark 6 telescope. We also consider the basic
performance characteristics of the Mark 6 telescope, and include the
results of a preliminary investigation of the telescope's energy
threshold.

\section{Large Area Flux Collectors}

\subsection{Overview}

The mirrors were designed be of adequate optical quality for use with
imaging cameras (pixel resolution $\approx 0.25^\circ$), lightweight, of
low cost and easy to produce in large numbers. A mirror focal length of
$7.0 {\rm~m}$ and an aperture of f/1.0 has been specified to give an
image scale compatible with a camera containing 91-elements of diameter
$0.25^{\circ}$ and allow the effective use of light collecting cones
employing no more than a single reflection. Our specification of mirror
diameter, focal length and pixel size ensures that the image of a
\v{C}erenkov flash will not be distorted by the finite depth-of-field.

Ray tracing and measurements have confirmed that the off-axis
performance of parabolic mirrors is adequate for our purposes. While
segmented mirrors can be designed which have superior off-axis
performance compared to an ideal parabolic mirror by reducing the
effects of coma (e.g. the Davies-Cotton design employed by the Whipple
10 m telescope --- see \citeauthor{Lewis1990}, \citeyear{Lewis1990}),
they have disadvantages for event triggering and energy threshold due to
their non-isochronous performance.

\subsection{Manufacture of Large Area Mirrors}

The construction principle employed follows that used to produce the
antenna sections for the UK millimetre-wave telescope by the staff of
Rutherford Appleton Laboratory (J. Hall, private communication) and
adapted by us to produce mirrors for a range of \v{C}erenkov telescopes
since 1985 (\citeauthor{mark3paper}, \citeyear{mark3paper},
\citeauthor{mark4paper}, \citeyear{mark4paper}). For ease of
manufacture, each mirror is made from 24 close fitting sectors. Each
sector is filled with aluminium honeycomb material. To give rigidity,
the back of the honeycomb is bonded to a dural backplate and surrounded
by a dural frame. The front reflecting surface is a sheet of anodised
aluminium which is vacuum formed onto a convex steel plug and bonded to
the front surface of the honeycomb. 

The aluminium honeycomb employed in our mirrors is Aeroweb 3003 which
has a cell size of 0.8 cm and a foil thickness of 0.06 mm. The thickness
of the honeycomb web is 50 mm. The adhesive used to bond the honeycomb
to the front and back sheets is Redux 420A/B (Ciba-Geigy). The material
used for the reflecting surface is Alanod 410G Special aluminium sheet,
of thickness 0.5 mm, which is anodised during manufacture to give a
specular reflection of $\geq 75\%$ over a wavelength range of 300 -- 700
nm. The rear surface of the mirror is dural sheet of 1.6 mm thickness.

During manufacture the mirrors are cold cured for 24 hours to ensure a
faithful reproduction of the shape of the plug. The mirrors formed using
this technique are identical; the limit to quality is the form of the
plug. After curing, the mirror segment is released from the plug and
trimmed to shape. Any gaps in the rear surfaces of the mirrors are then
filled with resin, and painted white to limit solar heating (which
occurs when the telescope is parked with its rear to the sun).

The weight of each mirror sector is 14 kg and the material cost is
\pounds 150. Manufacturing effort is $\sim 4$ technician hours per
sector, excluding the fabrication of the plug.

Further improvements to these mirrors are possible by reskinning with a
new Alanod reflective surface formed over a more accurate plug.

\subsection{Characteristics of Mirrors}

\subsubsection{Reflectivity}

The total reflectivity of the mirror surface material was measured to be
$ \geq 83\%$ over the wavelength range of interest to us ($\sim 300 -
500~{\rm nm}$) (T.C. Weekes, private communication; A.F. Vickers,
private communication). Specular reflection is measured to be $> 75\%$.
The reflectivity of a sample of the front surface material was not
significantly deteriorated by exposure to an industrially-polluted
atmosphere over a period of 24 months. No significant deterioration in
the mirrors of the Mark 3 telescope has been observed in over ten years'
exposure to the elements.

The requirement for accurate field alignment of three flux collectors,
each comprising 24 of these large sectors, necessitated development of a
new alignment procedure. A self contained, rotating laser system within
a 4 m length of rigid aluminium channel was produced, which provided a
system of parallel laser beams which could be swept over each mirror
segment in turn, imitating the light from a distant light source. The
spin axis of the laser system and the nominal optic axis of the flux
collector were adjusted until they were coincident. The principal laser
beam passed through the focus, which is at the centre of the detector
package at the focal plane, and impinged on the centre of the flux
collector. Once this condition was achieved the mirror sectors were
adjusted successively to bring their individual foci to the collective
focus of the flux collector.

\subsubsection{Image Formation}

\begin{figure} 
\centerline{\psfig{file=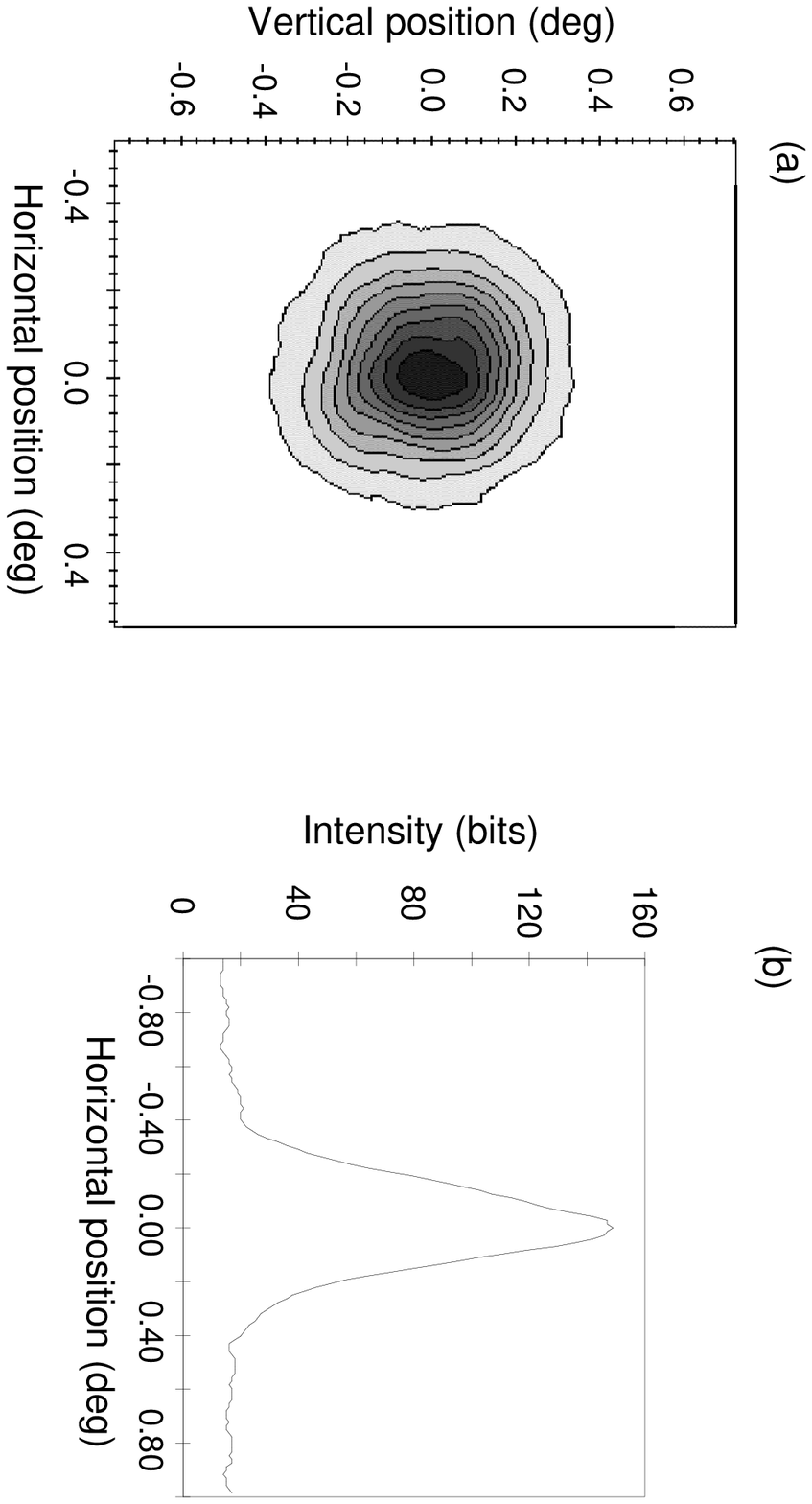,height=7cm,angle=90}}

\caption{(a) A stellar image produced by the central
mirror on the Mark 6 telescope and recorded with an optical CCD camera
viewing the image formed at the prime focus. (b) Horizontal sections
through the CCD image. The FWHM is about 36 mm, corresponding to an RMS
angular resolution of $0.18^\circ$.}\label{starimage} 

\end{figure}

The success of the fabrication and alignment of the mirrors has been
tested by focusing a stellar image onto the image plane and recording
this image with a CCD camera. In Figure~\ref{starimage} we show the
results of such a measurement. The RMS of the recorded image is $0.18 ^
\circ$. A cross-section through the image is also shown. The images
formed by all three collectors are similar, and have remained stable
over a period of 3 years.

\section{Telescope Mount and Steering}

\subsection{Alt-azimuth Mount}

The support for the Mark 6 telescope is a purpose made alt-azimuth
mount, the design of which has been influenced by those successfully
employed in our earlier telescopes.

The three flux collectors are rigidly connected and are supported on a
cradle by two bearings on a horizontal axis. The structure is
approximately in balance at all zenith angles.

\subsection{Telescope Steering/Pointing}
\label{steering_sect}

The control of the attitude of the telescope is via DC servomotors
driving onto gears mounted directly on the telescope structure. Both
motors incorporate an integral gearbox, while the azimuth drive has a
$40 - 200$ N m variable torque limiter to protect the motor and gearbox
against damage caused by wind forces on the telescope. Angles are sensed
by absolute digital 14-bit shaft encoders with a resolution of
$0.022^{\circ}$. The calculated azimuth and zenith of a source is
compared by a digital servomechanism (employing 12-bit resolution) with
the shaft encoder outputs at 100 ms intervals. The error signals are
passed via DACs to the DC motor amplifiers. These provide damping on
acceleration and stabilise the movements of the telescope structure.
Thus although the telescope pointing is known to a resolution of
$0.022^\circ$, the source can be offset from the camera centre by up to
$0.1^\circ$ by this mechanism. This offset is corrected for in the
calculation of the image parameters used for gamma-ray selection (see
\citeauthor{1706_paper} \citeyear{1706_paper}, \citeyear{cenx3_paper}).

\subsection{Direction Sensing/Measurement}
\label{direction_sect}

The attitude of the telescope is measured in two ways. The shaft encoder
positions are recorded for each event to 14-bit accuracy. This gives a
measurement of the telescope's pointing to $\pm 0.022^{\circ}$ within
the telescope. In addition, a coaxial optical CCD camera (Santa Barbara
Instrument Group ST-4 using a 50 mm $f/1.4$ lens) is mounted on the
telescope and is used to absolutely calibrate the shaft encoders on a
continuous basis. A typical integration time of 3 s is employed. The
output of this CCD camera is continuously monitored by microcomputer,
which measures the position and brightness of a nominated guide star
within the $ 2^\circ \times 2^\circ$ field. This information is
integrated into the data stream on an event-by-event basis. Guide stars
of magnitude $m_{\rm{v}} \leq 6$ can be employed, providing absolute
position sensing better than $0.008^{\circ}$.

With a typical CCD star field the absolute pointing position of the
telescope can be updated in a time of 3 s which can be compared with a
typical event rate of 15 Hz. Full details of the extraction of the
absolute position of the telescope will be presented elsewhere
\cite{telescope_pointing}.

\subsection{Pointing Accuracy}

False source analysis has been shown to be a useful method of
demonstrating that $\gamma$-ray like events originate from the source
direction \cite{Kifune1995}. Conversely, such analyses may also be used
to verify the steering performance of a telescope, if an established
$\gamma$-ray emitter is observed.

\begin{figure}
\centerline{\psfig{file=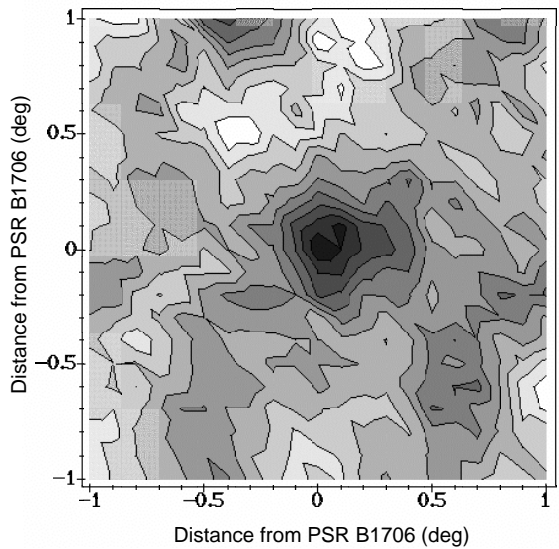,height=8cm}}

\caption{The pointing accuracy of the Mark 6 telescope demonstrated with
data taken during an observation of PSR B1706-44. Contour lines are in
$0.6~\sigma$ steps.}\label{Crab_point}

\end{figure}

An analysis of events from PSR B1706-44 selected as $\gamma$-ray
candidates has been made on the basis of shape alone
\cite{chadwick1706}, in which the position of the source has been
assumed to be at a matrix of positions in the field of view of the Mark
6 telescope's camera. For each false source position, the values of {\em
ALPHA} for the events have been calculated and the selection criterion
{\em ALPHA} $< 22.5^\circ$ applied. We show in Figure~\ref{Crab_point} a
plot of the significance of the source detection as a function of
assumed source position. The relative accuracy of the Mark 6 telescope
pointing is found to be $\leq 5$ arcmin.

\section{Photodetectors}

\subsection{Overview}

The Mark 6 telescope contains a camera consisting of 91 one inch
diameter PMTs surrounded by a guard ring of 18 two inch diameter PMTs
placed at the focus of the central flux collector. At the focus of each
of the left and right flux collectors is a detector package consisting
of 19 hexagonal PMTs (56 mm across the flats). All detector packages
have flux concentrators fitted; in the case of the camera these are
designed to increase the amount of light collected by the PMTs, whereas
in the left and right triggering detectors the flux concentrators
improve the temporal performance of the PMTs.

\begin{figure}
\centerline{\psfig{file=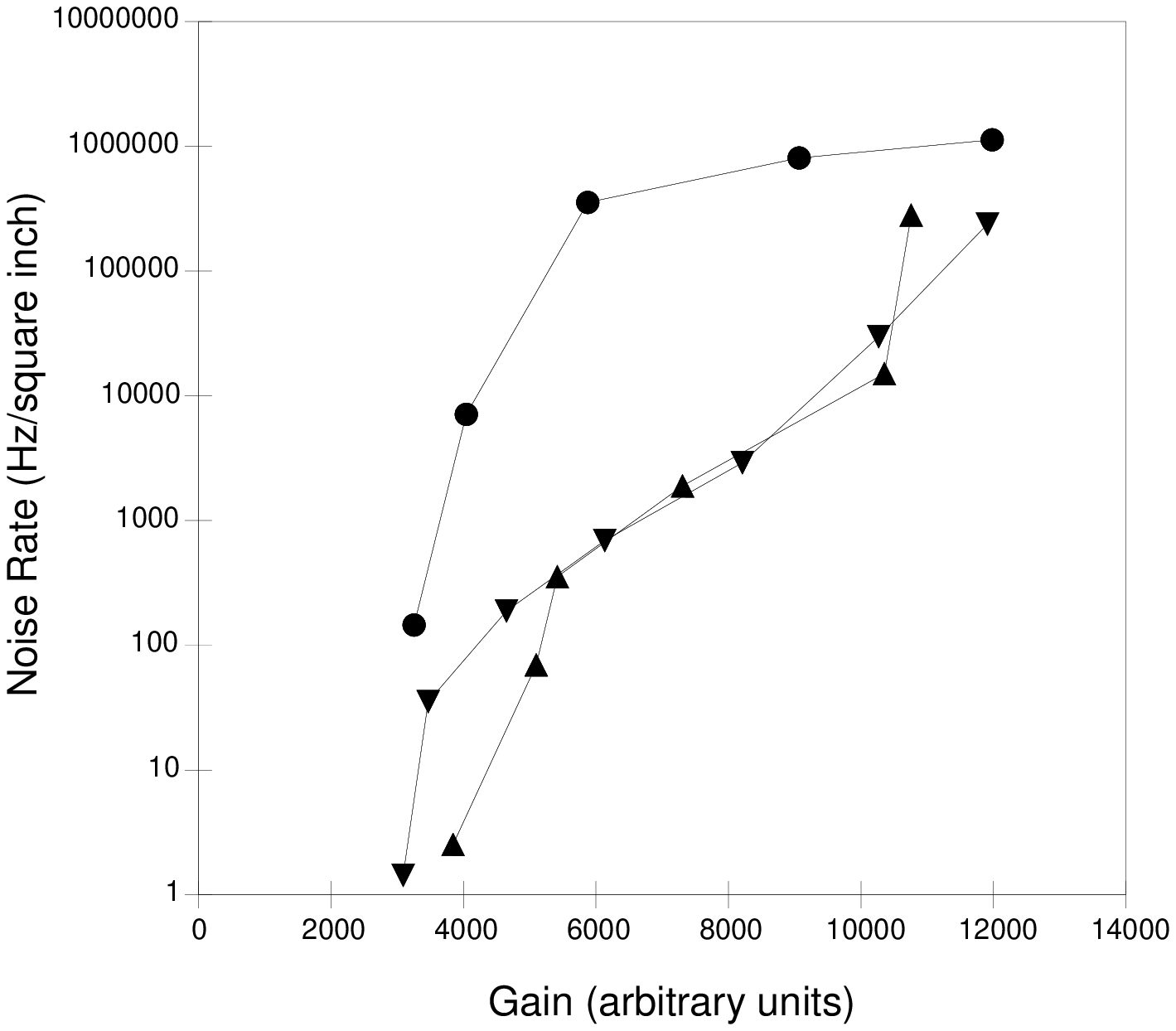,height=8cm}}

\caption{The variations of noise rate (normalised to unit photocathode
area) at constant illumination with gain of a typical Burle 8575 50 mm
diameter circular PMT (denoted by $\bullet$), a Hamamatsu R1924 25mm
circular PMT (denoted by $\blacktriangle$) and a Philips XP3422 55mm
hexagonal PMT (denoted by $\blacktriangledown$). In the Mark 6
telescope, the large triggering PMTs are operated at noise rates of
about 10 kHz. At these noise rates, the advantages of the Philips XP4322
over the Burle 8575 is apparent.}\label{PMTperform}

\end{figure}

\subsubsection{Hamamatsu R1924 $1''$ Diameter Circular PMT}

The performance of this tube is shown in Figure~\ref{PMTperform} where
the noise rate (for events exceeding a 50 mV discrimination level after
$\times 10$ amplification) is plotted versus the gain. The choice of PMT
for a high-resolution imaging camera was between this and the
corresponding Burle tube (type S83062E); the choice was determined
largely on the basis of ease of mechanical assembly, since the
performance of the two PMTs was similar.

\subsubsection{Phillips XP3422 55mm Hexagonal PMT}

\label{hex_section}

The Phillips XP3422 is an 8-stage hexagonal PMT, with 56 mm useful width
across the flat sides and a bi-alkali photocathode. It has not been used
previously in ground-based gamma-ray astronomy. It has a linear focussed
dynode structure with CuBe dynodes. The nominal risetime (10\% -- 90\%)
of the tube is $\approx 3$ ns and the FWHM is $\approx 5$ ns, with a
transit time of $\approx 37$ ns.

A potential limitation of PMTs with large area photocathodes is the
dependence of transit time on the point of incidence of the photon on
the photocathode. We have measured transit time relative to the response
of a Hamamatsu $1''$ PMT. The transit time for the Phillips XP3422 PMT
is at least 12 ns greater than that for the $1''$ PMT. Furthermore, the
variation in transit time between photons arriving on the central axis
and those arriving at the vertices of the PMT face is 10 ns. Although
the nominal timing characteristics of this tube are worse than those of
the 8575 $2''$ diameter and R1924 $1''$ diameter PMTs, the inevitable
degradation of the signal when transmitted down $\approx 50$ m of cable
from the telescope and the effect of the recording electronics limits
the importance of this disadvantage. The temporal performance provides
the motivation for the addition of flux concentrators to direct the
\v{C}erenkov light towards the centre of the photocathode. The flux
concentrators are described in Section~\ref{triggering_detectors}. 

A subset of a large sample of these tubes has been found to have an
unusually good ratio of the response to \v{C}erenkov light to the noise
--- see Figure~\ref{PMTperform}. Thus, although the inherent temporal
response of these hexagonal tubes is relatively poor, the greatly
improved noise performance and the use of light cones to enhance the
temporal performance results in these tubes being useful for experiments
aiming to detect low energy $\gamma$-rays.

\subsection{Triggering detectors}

\label{triggering_detectors}

\begin{figure}
\centerline{\psfig{file=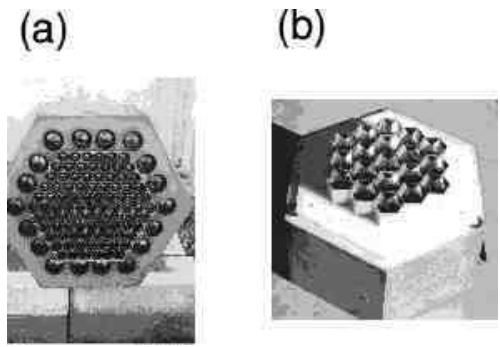,height=6cm}}

\caption{The prime focus detector packages employed in the University of
Durham Mark 6 gamma ray telescope. (a) shows the package deployed as the
central camera, while (b) shows the package deployed in the left and
right hand dishes.}\label{bullets}

\end{figure}

The photodetectors of the left and right flux collectors each consist of
19 Phillips XP3422 hexagonal tubes and are primarily used as a part of
the event selection system. The PMTs are arranged in a close packed
hexagonal pattern (see Figure~\ref{bullets}(b)). They are magnetically
shielded and electrically insulated, and are mounted with their shielded
hexagonal faces touching. The signal from the last dynode is taken from
the detector package to the control room via 50 m of $75\Omega$ coaxial
cable (type CT100). The applied voltages and gains of the PMTs are
individually adjustable.

We have added light concentrators to improve the temporal respose of
these PMTs. These are designed to concentrate the incident light on the
central 44 mm diameter circle of the hexagonal photocathode, and are
scaled-up versions of those developed for the $1''$ PMTs. While
employment of these light concentrators reduces the total amount of
light detected by the PMT by a few per cent, the rise time of the signal
reduces by 15\% and, most importantly, the height of the signal
increases by 15\% corresponding to a decrease in threshold of a similar
amount.

\subsection{Camera Detectors}
\label{highres_sec}

The detector package for the central collector of the camera consists of
91 $1''$ circular Hamamatsu R1924 PMTs arranged in a close-packed
hexagonal array, giving a pixel size of $0.25^{\circ}$, and surrounded
by 18 $2''$ circular Burle 8575 PMTs which form a guard ring. The
arrangement of tubes is shown in Figure~\ref{bullets}(a). Since the dead
area between the $1''$ tubes amounts to 45\%, we have constructed
conical reflective light guides for these tubes resulting in a 70\%
increase in the detected light pulse.

\section{Telescope Trigger}
\label{trigger_section}

The Mark 6 telescope uses a 3-fold spatial and 4-fold temporal
coincidence system. The event selection for the Mark 6 telescope
consists of two elements:

\begin{enumerate}

\item The left/right flux collectors, each viewed by 19 hexagonal PMTs.
An event selection is initiated whenever a signal is detected from a
pair of PMTs at similar positions in the left/right detectors.

\item The central flux collector. This is viewed by the `camera', which
includes 91 $1 ''$ PMTs which cover a similar area of the sky as the
left/right PMT matrices. For an event to be selected, it is required
that any 2 of the 7 camera PMTs which cover the same area of sky as the
left/right PMTs which have responded also produce a signal and that the
2 PMTs are adjacent.

\end{enumerate}

In the Mark 6 telescope, selections 1 and 2 above are both achieved
using a hardware first-level trigger with a decision time of $\sim 10$
ns (see Section \ref{logging_section}). The discriminator threshold is
set at 50 mV for all PMTs, the signals from which are first subject to
$\times 10$ amplification. The most sensitive element of the system at
present is the central flux collector/camera combination; work on
increasing the sensitivity of the left/right detectors continues.

\section{Electronic Systems}

\subsection{Overview}

The design of the recording and control electronics for the Mark 6
telescope has followed the pattern developed for our previous
telescopes. We separate the telescope performance monitoring and data
logging functions, and employ a network of distributed computers to
control and monitor the telescope's performance and log the data.

\begin{figure}
\centerline{\psfig{file=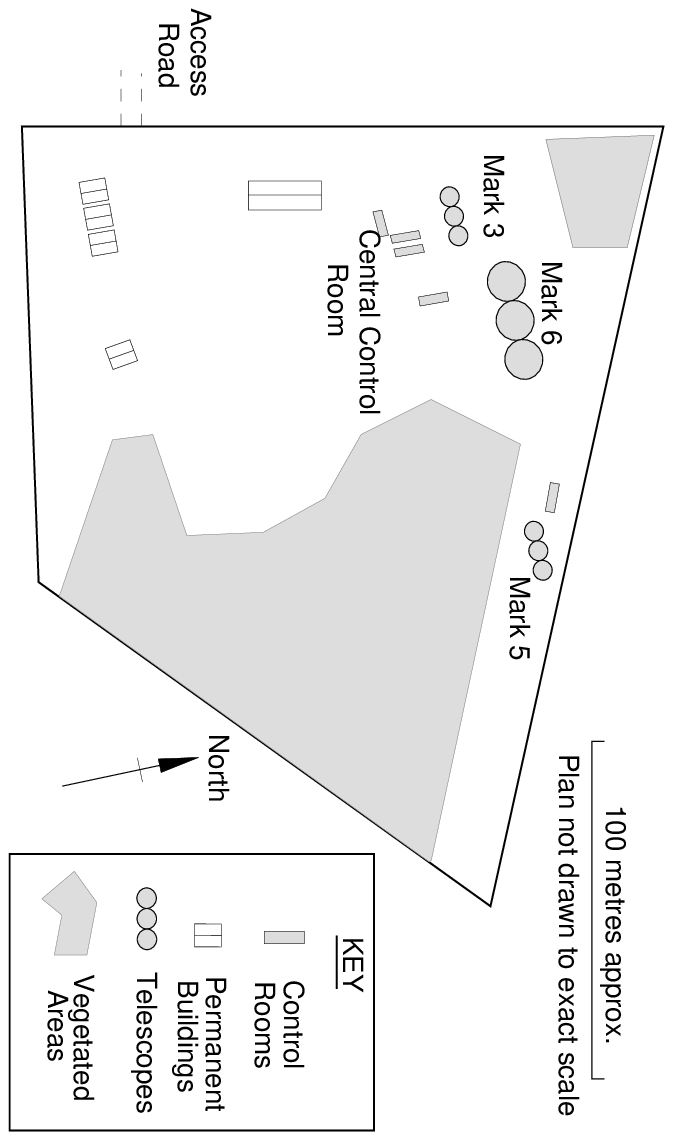,height=7cm,angle=90}}

\caption{The Narrabri telescope environment, showing the three
operational telescopes and the central control room.}\label{site}

\end{figure}

The Mark 6 telescope has been designed to be integrated into the
Narrabri telescope environment, where it operates in conjunction with
the existing telescopes. A schematic diagram of the telescope
environment is shown in Figure \ref{site}. Each telescope has a
self-contained control room adjacent to it, which houses the control and
logging electronics and computers. 

The central control room, from which the operation of all the telescopes
can be monitored, is near the Mark 3 telescope. The individual
telescopes' control rooms and the central control room are connected by
two local area networks (LANs), a medium bandwith LAN for control
purposes and a high bandwidth LAN for data transfer. Each telescope
passes performance information to a display computer in the central
control room so that its operation can be continuously monitored. Also
relayed to the central control room are the steering and logging
computer displays, and the display from the paraxial CCD camera, so that
telescope pointing and logging can be monitored.

A rubidium oscillator is used as the site-wide local time standard (see
Section~\ref{timekeeping_sect}). The output from this oscillator is used
by each telescope for timestamping the arrival of every \v{C}erenkov
flash. A microcomputer and associated electronics produce a randomly
occurring trigger pulse which is distributed to all telescopes and mixed
with their event trigger to provide a measure of the digitizers'
pedestals.

Event triggers from each telescope are passed via high bandwidth cable
to the Mark 5 telescope control room, where TDCs are used to measure the
arrival time differences between events which trigger two or more
telescopes on the site.

\subsection{Telescope Performance Monitoring System}
\label{TPMSsec}

The general performance and stability of the telescope is monitored by
measuring the anode currents and single-fold rates from each of the 147
PMTs, 3-fold event rates, ancillary data (mirror, laser and camera
temperatures, wind speed, {\em etc.}) and steering information. Anode
currents are monitored by purpose-designed ADC units, while the
single-fold rates are monitored using fast scalers. In addition, the
`2-of-7' rates from the central camera and the 3-fold coincidence rates
from each of the 19 triggering channels are measured and displayed.

\begin{figure}
\centerline{\psfig{file=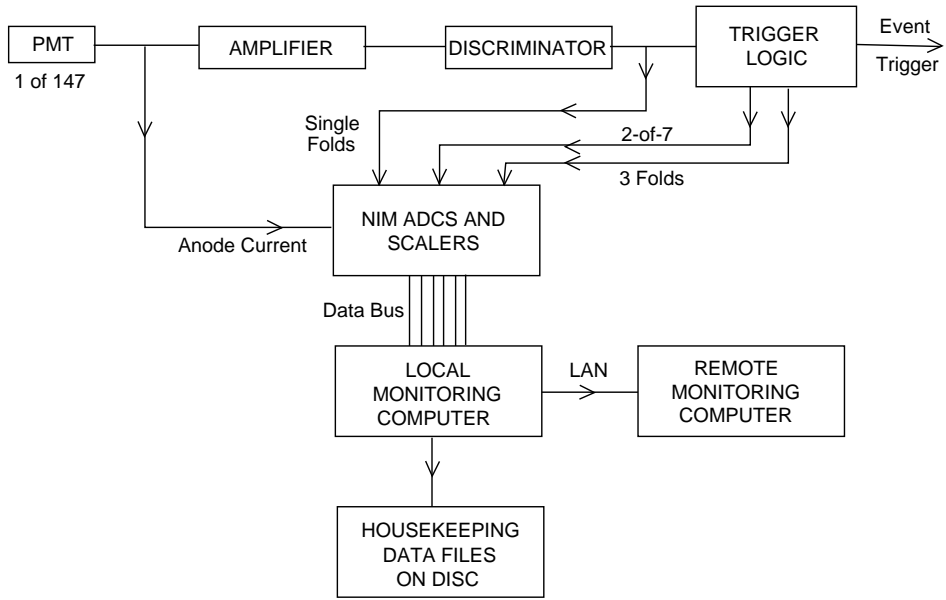,height=8cm,angle=90}}

\caption{The telescope performance monitoring system used in the Mark 6
telescope.}\label{TPMS}

\end{figure}

The ADCs and scalers are read out by a 32-bit RISC computer which, as
well as displaying the measured anode currents and single-fold and
coincidence rates, performs a series of checks that the various
parameters are within predefined limits and will produce audible and
visible warnings for the operator should any of the parameters fall
outside these limits. A schematic diagram of the telescope performance
monitoring system (TPMS) is shown in Figure~\ref{TPMS}. 

The telescope performance data are updated every 6 seconds, and this
information is then broadcast over the local network to the
corresponding remote monitoring computer in the central control room,
which has an identical set of display and warning options. A limited
subset of performance data is included in every event record; every
minute, the complete performance data are recorded for subsequent
off-line analysis.

\subsection{Data Logging System}
\label{logging_section}

\begin{figure}
\centerline{\psfig{file=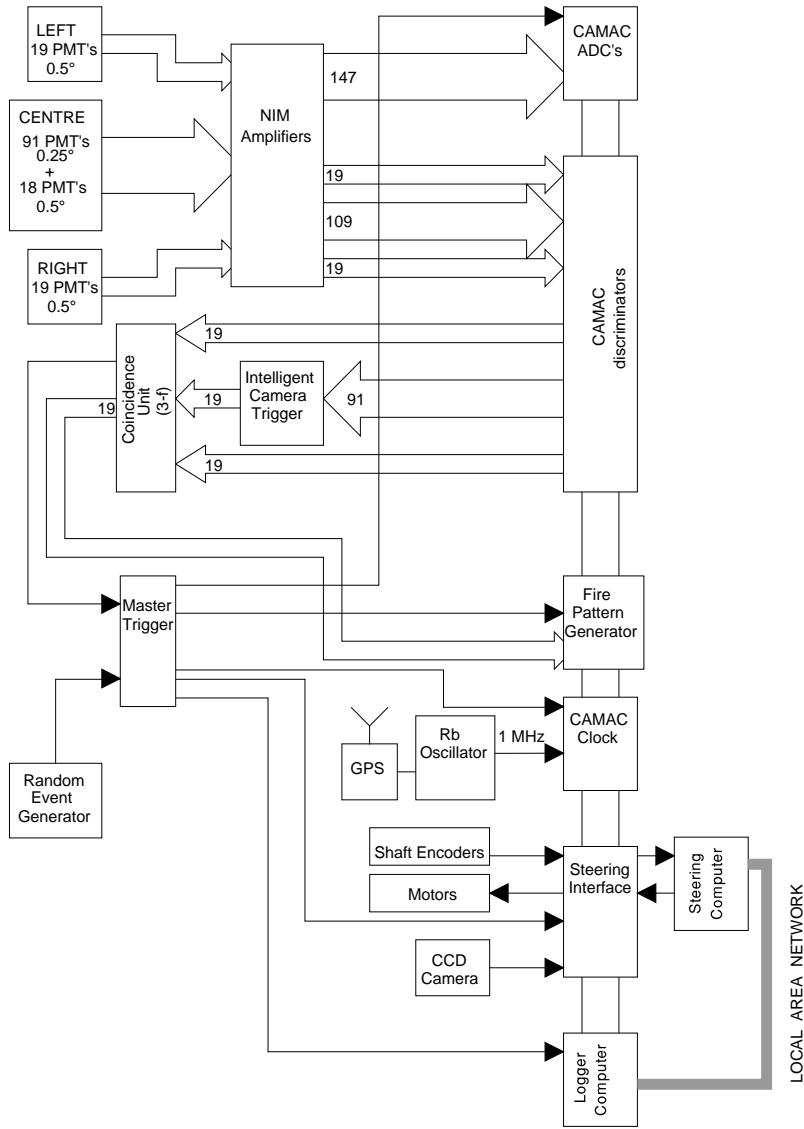,height=15cm}}

\caption{The data recording system employed in the Mark 6
telescope.}\label{electronics}

\end{figure}

A schematic diagram of the data logging system employed in the Mark 6
telescope is shown in Figure~\ref{electronics}. It is similar to that
used in our previous telescopes \cite{mark3paper}.

The signal from each PMT is passed down 50 m of CT100 air-cored cable
into the control room, where the signal transit time is trimmed using a
short length of RG179 cable to allow for individual PMT transit time
differences. The analogue signal then enters an automatic gain control
(AGC) unit which performs a number of functions:

\begin{enumerate}

\item the impedance of the signal path is changed from $75\Omega$ to
$50\Omega$ for subsequent processing

\item the anode current is sampled and converted to a voltage for
subsequent monitoring

\item the anode current may be compared with a preset value and an error
signal is generated to drive an LED within the PMT's field of view, so
that the PMT can be run under constant illumination conditions, if so
desired

\item the analogue signal is AC coupled and output for subsequent
processing.

\end{enumerate}

The output from the AGC unit is passed to a fast amplifier (LeCroy Model
612A) where it is amplified by a factor of 10 and fanned out. One output
from each channel is delayed by an appropriate amount and taken to the
input of a fast charge digitizer (LeCroy Model 2249A). The second output
from the amplifier is taken to the input of a fast discriminator (LeCroy
Model 4413). An output from each of the discriminators is used for
monitoring the noise rate of each PMT, while the other outputs are used
to form coincidences between a pair of outputs from the left and right
detectors and a corresponding response in the central camera. The
coincidence system has been designed to be flexible so that various
approaches to defining a camera trigger can be employed. The resulting
left, right and camera signals are passed to 3-fold coincidence units
(LeCroy Model 4516), with an effective coincidence resolving time of
$\leq 10$ ns.

The Mark 6 telescope is normally operated in a mode where the central
camera response is defined to occur when any two adjacent members of a
group of seven $1''$ PMTs (which view the same area of sky as an
individual left/right camera PMTs) have passed a preset discrimination
level. This coincidence requirement is implemented in hardware, using a
purpose-designed module employing ECL logic. The resolving time for this
`2-of-7' circuit is less than 10 ns.

Individual ECL outputs from the logic unit, corresponding to a trigger
in any of the 19 `3-fold' channels, are taken to a `coincidence
register', where the telescope `fire pattern' is latched for subsequent
recording. All event rates are scaled and recorded.

The OR'ed outputs from each section of the logic units are stretched and
taken to a voter coincidence unit where they are mixed with the pedestal
triggers to form a master trigger for the telescope. This master trigger
is then fanned out to provide (in the correct time sequence):

\begin{enumerate}

\item an interrupt to signal an event to the logging computer

\item signals to latch the coincidence registers, the telescope steering
information (both from the shaft encoders and the CCD star tracker) and
the event time

\item gate pulses for the fast charge digitizers

\item a signal for inter-telescope timing purposes.

\end{enumerate}

In addition, the logger system can record a limited amount of
information (event time and telescope `fire pattern') for
events occurring within the readout time of a preceding event.

\begin{figure}
\centerline{\psfig{file=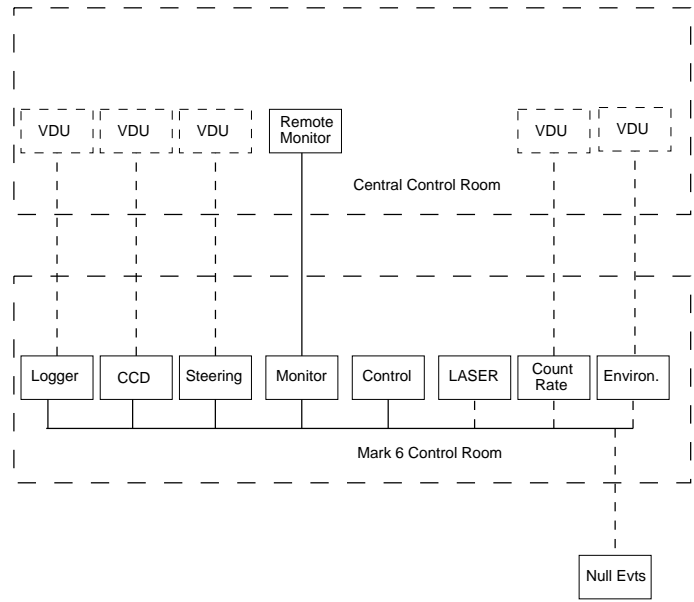,height=8cm}}

\caption{The system of computers used to control the Mark 6
telescope.}\label{computers}

\end{figure}

The network of computers employed to control and monitor the Mark 6
telescope is shown in Figure~\ref{computers}.

The logging computer is a 32-bit RISC computer, which communicates with
the CAMAC electronics via HyTec Model 1330 CAMAC interfaces. The logger
is interrupt driven and has a deadtime of 2.3 ms (essentially limited by
the readout time of the CAMAC crates). 0.5 kbyte of data is collected
per event and data are recorded on a local 2 Gb hard disk for subsequent
archiving.

The performance of the telescope is monitored using a 32-bit RISC
computer (see Section~\ref{TPMSsec}), which broadcasts the monitoring
information to the central control room. The CCD star tracker is also
controlled by a 32-bit RISC processor (see
Section~\ref{direction_sect}). The system used for steering the
telescope is described in Section~\ref{steering_sect}; it employs an
8-bit microcomputer using a system common to all telescopes.

All the computers operating the Mark 6 telescope communicate via a LAN
and all operator interaction during normal operation is via a single
controlling computer, which supervises all the tasks associated with
observing, including passing target coordinates to the steering computer
and commanding the steering computer to commence tracking the object and
controlling the logger and monitor computers.

\subsection{Timekeeping}
\label{timekeeping_sect}

\begin{figure}
\centerline{\psfig{file=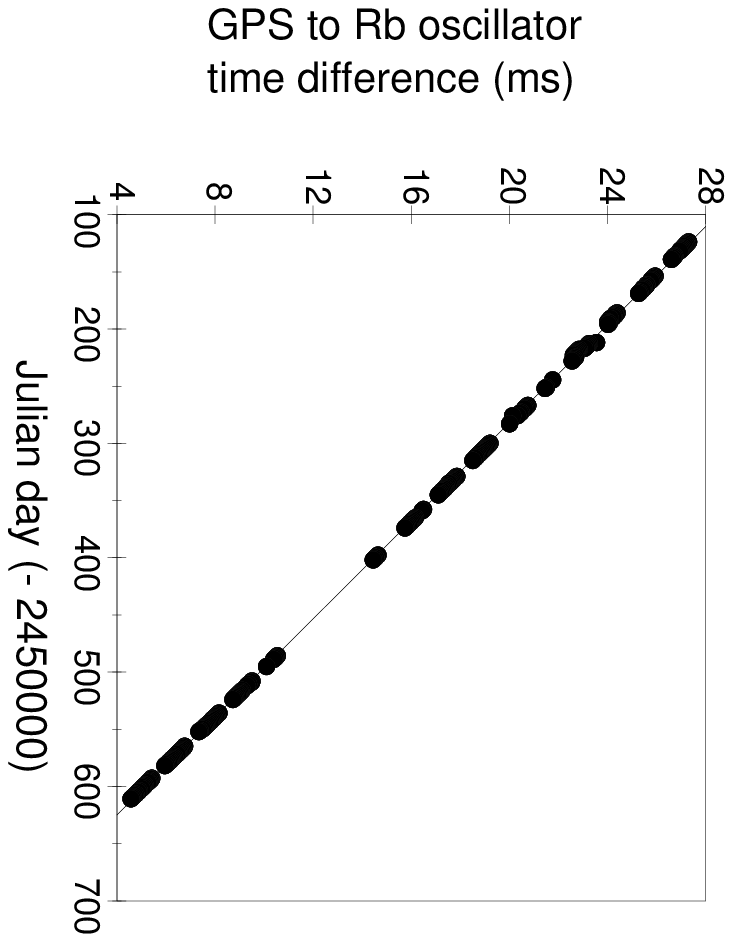,height=8cm,angle=90}}

\caption{The measured drift rate of the Rb oscillator relative to a GPS
timing signal over a period of 1 year.}\label{driftrate}

\end{figure}

A rubidium atomic oscillator provides a frequency standard for the whole
of the Narrabri site. We employ an Efratom Model FRK-L Rb oscillator,
which provides a stable 10 MHz reference signal. The drift rate is
regularly monitored and measured by comparison with the signal from a
GPS receiver. This provides a daily absolute comparison with our local
time standard. All events are time stamped to a relative accuracy of $1\
\mu {\rm s}$ and with an absolute accuracy of $\leq 10~\mu {\rm s}$.
In Figure~\ref{driftrate} we show the measured drift rate of the Rb
oscillator relative to GPS-derived time over a period of 1 year.

\section{Calibration}
\label{calibration}

In order to measure accurately the images from such a telescope it is
necessary to know the gains of each PMT, so that the digitized output
can be accurately related to the observed \v{C}erenkov photon density.

\subsection{Absolute Calibration}

Calibration of the PMTs and electronic channels is performed initially
by the use of a radioactive light pulser. This consists of $^{241}{\rm
Am}$ dispersed in a small piece of plastic scintillator, and produces an
optical flash of $\approx 3$ ns duration at a rate of about 1 kHz. The
light pulser is placed at a fixed distance from the front face of each
PMT in turn (in complete darkness) and the pulse area spectrum produced
is recorded. Knowledge of the pulse area spectrum of each tube and the
emission characteristics of the light pulser enable the absolute gains
of the tubes to be calculated.

The light output of the pulser is, according to the manufacturers'
specifications, 300 photons per pulse. Independent measurements indicate
an output of $\sim$ 70 $\pm$ 7 photoelectrons per pulse, depending on
the quantum efficiency of the PMT responsible for the measurement.

\subsection{Relative Calibration}

The absolute gain measurement method described above suffers from two
disadvantages:-

\begin{enumerate}

\item the amount of light emitted by the light pulser is so small that
the calibration can only be done with the PMT in the dark for the signal
to be detected above the noise

\item calibration cannot be performed continuously throughout an
observation.

\end{enumerate}

To circumvent these problems, a method has been developed using a
nitrogen laser to provide a diffuse source of blue light from a plastic
scintillator. By arranging for the telescope to be triggered by a flash
of laser induced light, gain calibration measurements can be
incorporated directly into the datastream and the telescope triggering
monitored. However, relative gain calibration only is possible, since
the number of photons emitted per laser flash varies and depends on, for
example, the laser temperature.

The 3 ns, 40 kW pulse of 337 nm radiation from a nitrogen laser (VSL
Model 33700) impinges on a 5 cm cube of plastic scintillator (NE102),
which converts the laser pulse to a pulse of $400 \pm 20\ {\rm nm}$
radiation. This blue light is transmitted by means by plastic fibre
optic cables to the centre of each of the three mirrors. There is little
dispersion suffered by the pulse along the 20 m of fibre optic cable.
The output from the fibre optic cable at the centre of each mirror is
diffused using a flashed opal diffuser, producing a pulse of light which
is uniform over the face of the detector package.

The lengths of the left, right and centre fibre optic cables are trimmed
to enable the respective calibration pulses to trigger the telescope. A
single PMT is placed in a collimator next to the camera such that it
views only the light pulse produced by the laser and is shielded from
light reflected off the mirror. Signals from this PMT then provide a
simple means of reliably identifying and monitoring the calibration
pulses, which are recorded in an identical manner to real events. The
laser produces individual pulses separated by random intervals with a
mean rate of 50 min$^{-1}$ maintained throughout observations. This
procedure allows the PMT gain to be measured to $\pm$ 1 -- 2\% for each
15 minute data segment.

\subsection{Stability of Response}

\subsubsection{Introduction}

The Mark 6 telescope is equipped with comprehensive monitoring
equipment. Throughout observations, the performance of the PMTs is
measured. The PMTs' anode currents and noise rates are recorded using
the TPMS and their relative gains are measured. In addition, we monitor
environmental conditions such as wind speed, atmospheric pressure, air
temperature and the temperatures inside the detector packages and laser
system.

\subsubsection{Temperature Effects on PMTs}

Continuous monitoring of the performance of the Mark 6 PMTs during
observations have shown that their gains are affected by temperature
variations. Tests under controlled laboratory conditions confirm that
the gains of the PMTs typically fall by up to $1\% \pm 0.5\%$ for the
Phillips XP3422 and $0.5\% \pm 0.25\%$ for the Hamamatsu R1924 for every
$1^{\circ}$C rise. Power dissipated in the PMT dynode chains means that
the temperature in the camera rises above ambient in the first hour of
an observation, with a consequent change in PMT performance. To ensure
PMT stability throughout observations, all three detectors on the Mark 6
Telescope are equipped with temperature control systems. Heaters in the
detectors raise their internal temperatures before observations begin,
and thereafter their temperatures are thermostatically controlled. The
increase in PMT noise that this produces is small compared with night
sky noise.

\subsubsection{Geomagnetic Effects}

The PMTs used in the Mark 6 telescope are equipped with mu-metal
shielding. This has been shown to protect them from the effects of the
earth's magnetic field in laboratory and field tests. The change in gain
for a 0.5 gauss change in magnetic field is $< 1\%$ for a bare tube; for
a tube shielded with mu-metal we have been unable to measure an effect.

\subsubsection{Laser Stability}

Monitoring the laser from which we derive the relative gain calibration
pulses for the Mark 6 telescope suggested that the size of the laser
pulse not only varied from pulse to pulse but also changed
systematically with temperature. Measurements made in the laboratory
confirmed this was the case; the height of the pulse from the laser
decreases by less than $2.5\% \pm 0.5\%$ per $^\circ$C increase. The
temperature of the laser is measured routinely during telescope
operation.

\section{Performance of the Telescope}

The Mark 6 telescope has been operational since March 1995 and the
performance is summarized below.

\subsection{Response of Telescope}

\subsubsection{Cosmic Ray Count Rate}

\begin{figure}
\centerline{\psfig{file=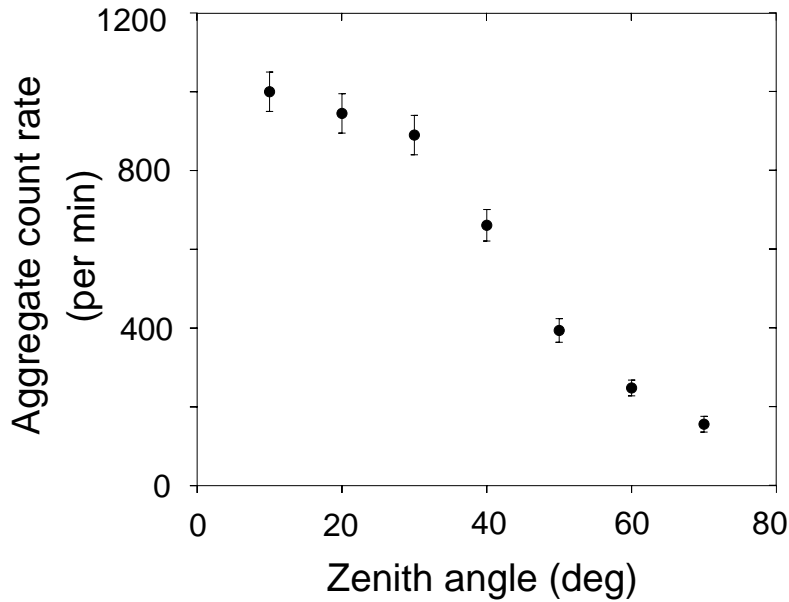,height=8cm,angle=90}}

\caption{The measured total telescope counting rate as a function of
zenith angle for the University of Durham Mark 6 telescope during normal
operation, employing a 3-fold spatial and 4-fold temporal coincidence
system (see Section \ref{trigger_section}).}\label{zenith}

\end{figure}

Count rates of up to 1000 per minute from a field of view of 9 square
degrees have been achieved with the Mark 6 telescope pointed near the
zenith. Figure~\ref{zenith} shows the variation of the count rate with
zenith angle. The measured zenith angle distribution is typical of that
observed with a multiple-mirror telescope system --- the Mark 6 trigger
is dominated by the medium-resolution left/right cameras.

\subsubsection{Event Size Distribution}

\begin{figure}
\centerline{\psfig{file=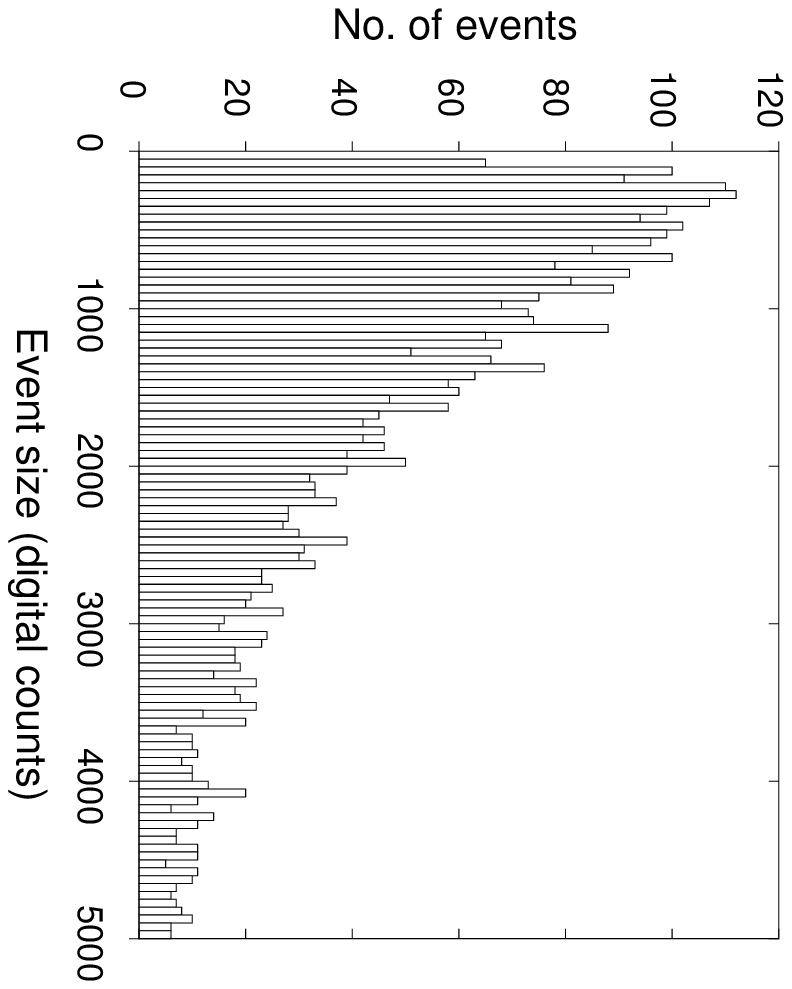,height=8cm,angle=90}}

\caption{The measured size distribution of events recorded by the Mark 6
Telescope.}\label{sizes}

\end{figure}

The distribution of \v{C}erenkov light flashes with event sizes
expressed in terms of digital counts is shown in Figure~\ref{sizes} (200
digital counts is equivalent to 125 GeV of energy for a $\gamma$-ray
event).

\subsubsection{Typical High Resolution Images}

\begin{figure}
\centerline{\psfig{file=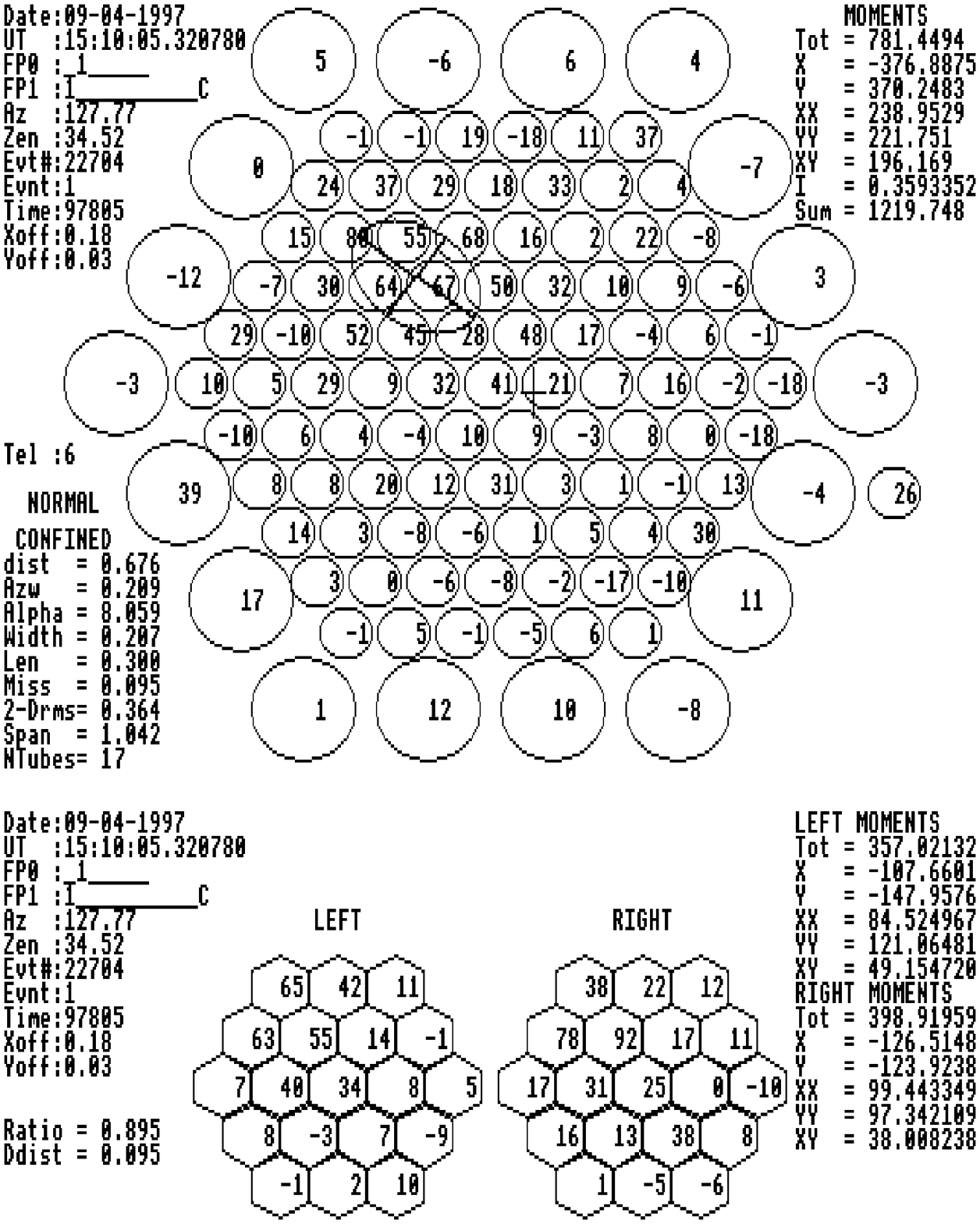,height=16cm}}

\caption{A typical image recorded with the Mark 6 telescope. Shown are
the image recorded with the central camera and the two images recorded
with the left and right detectors. The left and right detectors cover
the same area of the sky as the central 91 PMTs of the central
camera.}\label{camera_event}

\end{figure}

In Figure~\ref{camera_event} we show some typical data recorded with the
Mark 6 telescope. The response of each PMT in the camera has been
adjusted to allow for variations in digitizer pedestal and PMT gain.
Initial analysis of the camera images has followed the procedure adopted
by the Whipple collaboration \cite{feganetal}. Each PMT response is
labelled as (i) an `image' PMT corresponding to a signal of at least
37.5\% of the peak pixel, (ii) a `border' PMT corresponding to a signal
of at least 17.5\% of the peak pixel if it is next to an `image' tube or
(iii) neither of these when the response is set to zero. Categories (i)
and (ii) are used to calculate the shape parameters \cite{Hillas1985}
which describe the shape and orientation of the detected image.

\begin{figure}
\centerline{\psfig{file=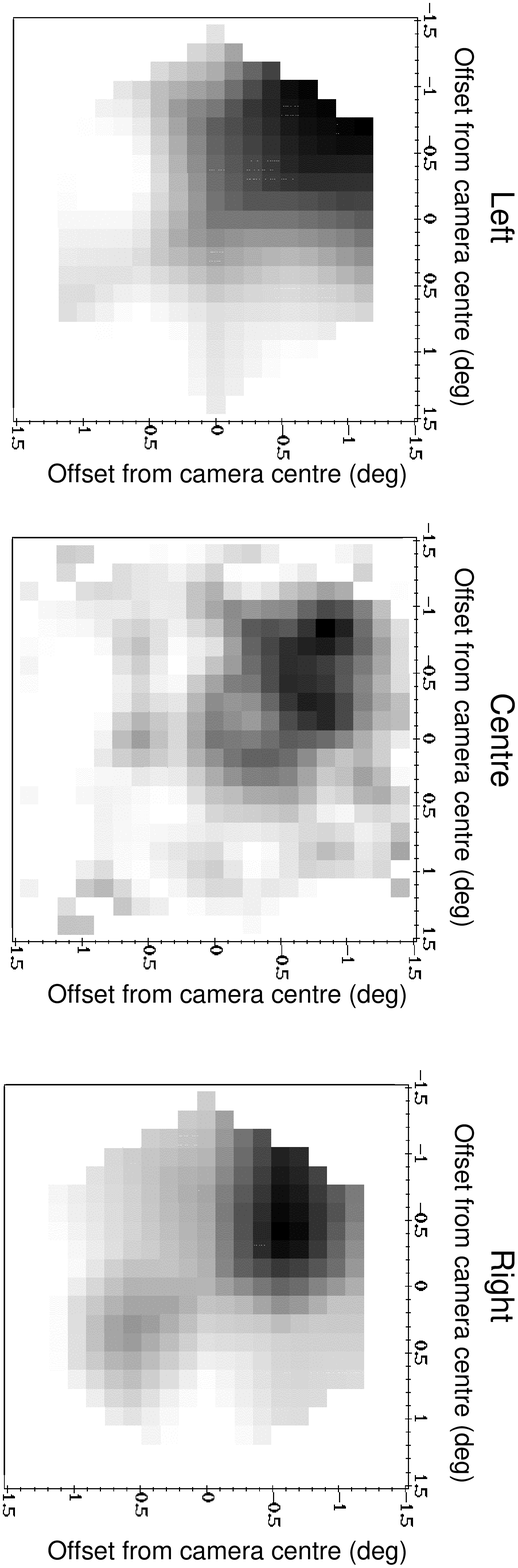,height=4.5cm,angle=90}}

\caption{A typical image recorded with the Mark 6 telescope. Shown are
the image recorded with the central camera and the two lower-resolution
images recorded with the left and right detectors. The event is the same
as that in Figure~\ref{camera_event} but is shown after interpolation
and reducing the left, centre and right images to a common
display.}\label{maple_event}

\end{figure}

\subsubsection{The Responses of the Triggering Detectors}

In addition to the high-resolution camera image, a 19-tube image is
obtained from each of the left and right detectors. This arrangement is
central to achieving a stable and low energy threshold. These detectors
also provide additional information on the \v{C}erenkov light and
further background rejection criteria. In Figure~\ref{maple_event} we
show the effect of interpolation and reducing to a common display the
data for the 3 imaging systems.

\subsubsection{Background Rejection Using Data from the Triggering
Detectors}

Simulations suggest that the characteristic distance for fluctuations in
the \v{C}erenkov light produced by a primary cosmic ray proton is about
15 m. The lateral distribution of \v{C}erenkov light from a VHE
$\gamma$-ray is much smoother than this. The triggering detectors on the
Mark 6 telescope are 15 m apart. Measurements of the fluctuations in the
\v{C}erenkov light from a shower made over this baseline can therefore
be used as a method of rejecting hadronic showers. Two simple measures
of fluctuation have been considered. The first is the ratio of the size
of the smaller to the larger of two light samples detected ({\em LRR}).
The size of the left and right samples are obtained by summing the
responses of each of the 19 pixels in each camera. It was expected that
$\gamma$-ray shower images should have values of {\em LRR\/} close to
unity. Simulations and results from observations agree that this
quantity does not produce a very efficient criterion for rejecting
background events. More effective is the angular distance between the
centroids of the two images ($D{\rm _{dist}}$). Simulations indicate
that selecting small values of $D{\rm _{dist}}$ produces a selection
with a Q factor of 1.4. This parameter has been used as part of the
routine analysis and appears to be as successful as the simulations
suggested \cite{chadwick97}.

\subsection{Energy Threshold}

\begin{figure}
\centerline{\psfig{file=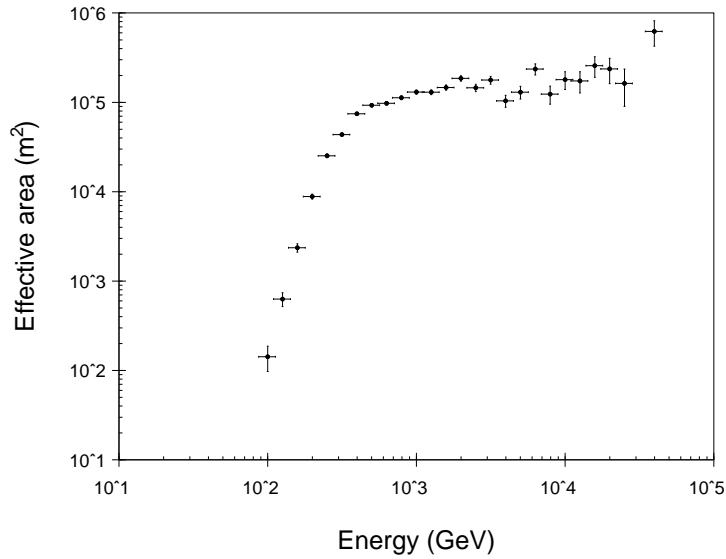,height=8cm}}

\caption{The effective area for gamma ray detection calculated for the
Mark 6 telescope}\label{effarea}

\end{figure}

\begin{figure}
\centerline{\psfig{file=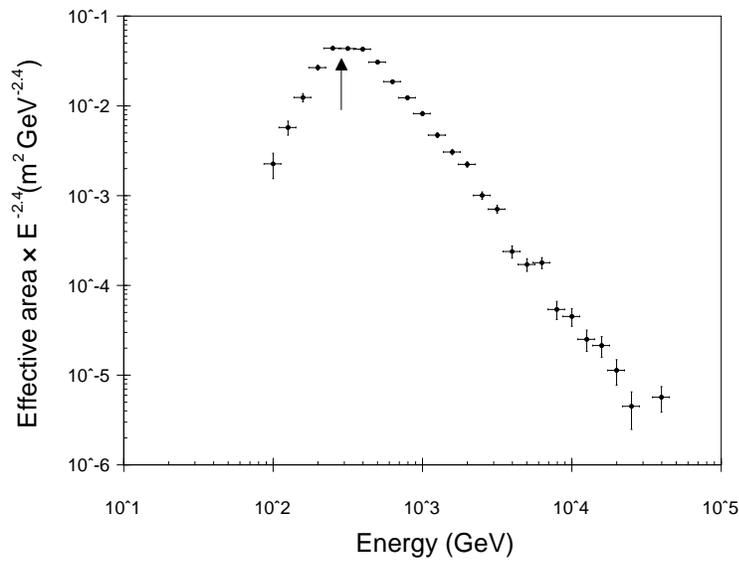,height=8cm}}

\caption{The simulated differential detection rate for gamma rays as a
function of energy threshold for the Mark 6 telescope. Also shown is the
input proton spectrum.}\label{turnover}

\end{figure}

The estimation of the energy threshold for the detection of gamma rays
by an atmospheric \v{C}erenkov telescope is a difficult task
\cite{Kohleretal}. The following is a preliminary approach to the
problem of estimating the energy threshold of the Mark 6 telescope.

Monte Carlo simulations of the photon yield from hadronic air showers
have been generated using the MOCCA program \cite{Hillas1982}. 40000
cosmic rays were generated from a circular field of view $2^\circ$ in
radius and out to a maximum impact parameter of 250 m. The simulations
were performed for a telescope inclined at $20^\circ$ to the zenith.
These were then presented to a model of the Mark 6 telescope, and by
altering the discriminator thresholds of the telescope model, the
trigger rate for the simulations was matched to the measured cosmic ray
count rate. Ideally, the telescope model should be further validated by
also matching the parameter distributions observed, but the preliminary
investigation described here has not included this.

This model of telescope performance was then applied to results from
Monte Carlo simulations of gamma ray air showers in order to estimate
the expected threshold energy for the detection of photons. 50000 gamma
ray showers were generated with energies ranging from 100 to $10^{5}$
GeV for a source with a power law spectrum of index $-2.4$. This is the
same as the measured value of the differential spectral index of the
Crab nebula in the VHE range \cite{Vacetal1991}. 

The effective detection area for a ground-based atmospheric \v{C}erenkov
telescope depends on the gamma ray energy, altitude, the \v{C}erenkov
light pool size and the triggering probability. Figure~\ref{effarea}
shows the effective detection area for the Mark 6 telescope as a
function of energy for gamma rays from a point source.

The threshold energy of such an atmospheric \v{C}erenkov telescope may
be defined in many ways. Here we use the energy at which the
differential gamma ray flux is a maximum. This is illustrated in
Figure~\ref{turnover}, which predicts a threshold energy for gamma rays
of $\sim 250$ GeV. There are many potential sources of error in this
estimate of energy threshold. For instance, the values of spectral
index, trigger conditions and mirror reflectivity applied will all
introduce uncertainties. Consequently, the systematic error in the
determination of the energy threshold is conservatively estimated as
$\pm\ 50\%$. It should also be noted that the values derived here are
the threshold energy and mean effective area assuming 100\% gamma ray
retention. Any background discrimination technique inevitably also
rejects a fraction of the gamma ray signal, altering the energy
threshold and effective area for gamma ray detection.

\section{Verification of Performance}

The performance of the Mark 6 telescope has been verified using data
from observations of PSR B1706-44 \cite{chadwick1706} and Cen X-3
\cite{cenx3_paper}. These have shown that, for events corresponding to
$\gamma$-ray energy $\ge~250$ GeV, conventional imaging techniques
enhance the $\gamma$-ray signal and provide a $6\ \sigma$ detection from
10 hours of data, after rejection of $> 99\%$ of background protons and
retention of $\sim 20\%$ of the $\gamma$-rays. For our observations of
PSR B1706-44 we find a Q-factor of $5.2$. The $\gamma$-ray source is
located to an accuracy of $\pm$ 5 arcminutes.

\section{Conclusions}

A new large VHE gamma ray telescope (the University of Durham Mark 6
telescope) was deployed on the Narrabri site in July 1994 and has
collected data since March 1995. In its present configuration, the
telescope has a gamma ray energy threshold of 150 GeV.
Further reductions in the energy threshold will be possible.

The telescope is equipped with a 109-element camera in the central flux
collector. Imaging can provide effective discrimination between
gamma-ray and hadron initiated events for gamma ray energies above 300
GeV.

Other papers in preparation will address the analysis techniques applied
to the Mark 6 telescope data, including the use of the information from
the left and right detectors and the low energy performance of the
telescope. 

\acknowledgements{}

This work was funded by the U.K. Particle Physics and Astronomy Research
Council. We are grateful to the University of Sydney for the lease of
the site at Narrabri. The assistance of the staff of the Main Workshop,
the Student Workshop and the Electronics Workshop of the Physics
Department, University of Durham, in the construction of the Mark 6
telescope is gratefully acknowledged.


\def\calgary_workshop{: 1993, in R.C. Lamb (ed.), {\em Towards a Major
Atmospheric Cherenkov Detector - II\rm}, Iowa State University: Ames,
pp. }

\end{document}